\documentclass[letterpaper, 10 pt, conference]{ieeeconf}

\IEEEoverridecommandlockouts        
\usepackage{cite}
\usepackage{amsmath,amssymb,amsfonts}
\usepackage{algorithmic}
\usepackage{hyperref}
\usepackage{epstopdf}
\usepackage{textcomp}
\usepackage{graphicx,color}
\usepackage{mathrsfs}
\usepackage[vlined,ruled]{algorithm2e}
\usepackage{subfigure}
\usepackage{url}
\usepackage{color}
\usepackage{dsfont}
\usepackage{bbm}
\usepackage{booktabs}
\usepackage{array}
\usepackage[dvipsnames]{xcolor}
\usepackage{yfonts}
\usepackage{cleveref} 
\usepackage{tikz,calc,bbding}\usetikzlibrary{positioning,shapes.geometric}
\usepackage[overload]{empheq}

\def\setsstar{(0.3, 0) ellipse [x radius=1.8cm, y radius=1.5cm, rotate=90]}

\def\imaB{(0, 0.2cm) ellipse [x radius=1.cm, y radius=1.cm,rotate=90]}
\def\setvstar{(-1.6, 0) ellipse [x radius=1.5cm, y radius=1.5cm,rotate=90]}
\def\setvstari{(-1.6, 0) ellipse [x radius=0.8cm, y radius=1.0cm,rotate=90]}
\def\setkerC{(-1.7, 0.2) ellipse [x radius=2.cm, y radius=2cm, rotate=90]}
\def\bounding{(-5,-2) rectangle (3.5,2.5)}

\usepackage{times} 
\usetikzlibrary{decorations.pathmorphing}

\newtheorem{theorem}{Theorem}[section]
\newtheorem{lemma}[theorem]{Lemma}
\newtheorem{definition}{Definition}

\newtheorem{remark}{Remark}
\newtheorem{assumption}[theorem]{Assumption}

\newcommand{\subscr}[2]{{#1}_{#2}}

\newcommand{\until}[1]{\{1,\dots,#1\}}
\newcommand{\fromto}[2]{\{#1,\dots,#2\}}

\newcommand{\Ker}{\operatorname{Ker}}

\newcommand{\rank}{\operatorname{rank}}
\newcommand{\Image}{\operatorname{Im}}

\newcommand{\real}{\mathbb{R}}
\newcommand{\complex}{\mathbb{C}}

\newcommand{\Basis}{\operatorname{Basis}}

\newcommand{\mc}{\mathcal}

\newcommand{\Bc}{\mathcal{B}}
\newcommand{\Cc}{\mathcal{C}}

\newcommand{\Rc}{\mathcal{R}}
\newcommand{\Sc}{\mathcal{S}}

\newcommand{\Vc}{\mathcal{V}}

\newcommand\Vstar{\Vc^*}
\newcommand\Sstar{\Sc^*}
\newcommand\Rstar{\Rc^*}

\DeclareSymbolFont{bbold}{U}{bbold}{m}{n}
\DeclareSymbolFontAlphabet{\mathbbold}{bbold}

\newcommand\oprocendsymbol{\hbox{$\square$}}
\newcommand\oprocend{\relax\ifmmode\else\unskip\hfill\fi\oprocendsymbol}

\renewcommand{\theenumi}{(\roman{enumi})}

\newcommand*{\QEDA}{\hfill\ensuremath{\blacksquare}}%
\newenvironment{pfof}[1]{\vspace{1ex}\noindent{\itshape Proof of
    #1:}\hspace{0.5em}}{
    }

\makeatletter

\SetKwInput{KwInput}{Input}                
\SetKwInput{KwOutput}{Output}              

\SetKwInput{KwInput}{Input}                
\SetKwInput{KwOutput}{Output}              
\SetKwInput{KwTransmit}{Transmit}   
\SetKwInput{KwReceive}{Receive}  
\SetKwInput{KwReturn}{Return}

\graphicspath{{img/}} 

\colorlet{red_fill_node}{Bittersweet!30}
\colorlet{green_fill_node}{OliveGreen!30}
\colorlet{blue_fill_node}{CadetBlue!30}

\colorlet{red_fill_net}{Bittersweet!10}
\colorlet{green_fill_net}{OliveGreen!10}
\colorlet{blue_fill_net}{CadetBlue!10}

\colorlet{tred}{Bittersweet}
\colorlet{tgreen}{OliveGreen}
\colorlet{tblue}{CadetBlue}

\makeatletter
\def\endthebibliography{%
  \def\@noitemerr{\@latex@warning{Empty `thebibliography' environment}}%
  \endlist
}
\makeatother

\begin{document}
\title{\bf Data-driven Meets Geometric Control: Zero Dynamics,
  Subspace Stabilization, and Malicious Attacks}

\author{Federico Celi  and Fabio Pasqualetti
  \thanks{This material is based upon work supported in part by
    awards AFOSR FA9550-20-1-0140 and FA9550-19-1-0235, and ARO
    W911NF2020267(NC4). Federico Celi and Fabio Pasqualetti are
    with the Department of Mechanical Engineering, University of
    California at Riverside,
    \{\href{mailto:fceli@engr.ucr.edu}{\texttt{fceli}},
    \href{mailto:fabiopas@engr.ucr.edu}{\texttt{fabiopas\}@engr.ucr.edu.}}}}
\maketitle

%

\begin{abstract} 
  Studying structural properties of linear dynamical systems through
  invariant subspaces is one of the key contributions of the geometric
  approach to system theory. In general, a model of the dynamics is
  required in order to compute the invariant subspaces of interest. In
  this paper we overcome this limitation by finding data-driven
  formulas for some of the foundational tools of geometric control.
  In particular, for an unknown linear system, we show how controlled
  and conditioned invariant subspaces can be found directly from
  experimental data. We use our formulas and approach to (i) find a
  feedback gain that confines the system state within a desired
  subspace, (ii) compute the invariant zeros of the unknown system,
  and (iii) design attacks that remain undetectable.
\end{abstract}

\section{Introduction}\label{sec:introduction}
The geometric approach is a collection of notions and algorithms for
the analysis and control of dynamical systems. Differently from the
classic methods in the frequency and state space domains
\cite{KJA-RMM:10,TK:80}, the geometric approach offers an intuitive
and coordinate-free analysis of the properties of dynamical systems in
terms of appropriately defined subspaces, and synthesis algorithms
based on subspace operations, such as sum, intersection, and
orthogonal complementation. The geometric approach has been
successfully used to solve a variety of complex control and estimation
problems; we refer the interested reader to
\cite{GB-GM:91,WMW:85,HLT-AS-MH:01} for a detailed treatment of the
main geometric control notions and their applications.

Similarly to the frequency and state-space approaches to control, the
geometric approach assumes an accurate, in fact exact, representation
of the system dynamics. To overcome this limitation and in response to
an ever-increasing availability of sensors, historical data, and
machine learning algorithms, the behavioral approach, and more
generally a data-driven approach, has seen a rapid increase in
popularity. Here, system analysis and control synthesis do not require
a model of the dynamics and are instead obtained directly from
experimental data reflecting the system dynamics~\cite{IM-PR:08}.

While analysis, control and estimation problems can often be solved
equivalently using different methods, the frequency, state-space,
geometric, and data-driven approaches all offer complementary insights
into the structure and properties of the system dynamics, and together
contribute to forming a comprehensive theory of systems. In this paper
we create the first connections between the geometric and data-driven
approach to system analysis and control. In particular, we derive
data-driven expressions of the fundamental sets used in the geometric
approach to solve a variety of control and estimation problems, and
show how these sets have an even more insightful and straightforward
interpretation when analyzed in the higher-dimensional data space as
compared to their geometric view in the lower-dimensional state space.

\smallskip
\noindent
\textbf{Related work.} From the seminal work \cite{GB-GM:69} that
introduced the notions of controlled and conditioned invariants, the
geometric approach to control has evolved over the last decades into a
full theory and a set of algorithms for linear
\cite{GB-GM:91,WMW:85,HLT-AS-MH:01} and nonlinear \cite{AI:95}
systems. Notable applications of the geometric approach are the
disturbance decoupling \cite{WMW-ASM:70} and fault detection
\cite{MAM-GCV-ASW:89} problems, the characterization of stealthy
attacks in cyber-physical systems \cite{FP-FD-FB:10y}, and the secure
state estimation problem \cite{HF-PT-SD:14}. In this paper we follow
the notation and techniques of \cite{GB-GM:91}, which we briefly
recall in Section \ref{sec: setup}.

The data-driven approach to system analysis and control is receiving
renewed and increased interest. While traditional indirect data-driven
methods use data to identify a model of the system \cite{MG:05} and
proceed to synthesize a controller in a second step, direct
data-driven methods bypass (at least apparently
\cite{VK-FP:21,FD-JC-IM:21,HJW-JE-HLT-MKC:19}) the identification step and
design
control actions directly from data. In this framework, recent results
tackle various problems for linear systems, including optimal
\cite{GB-DSB-FP:20,NM:20}, robust
\cite{CDP-PT:19,AB-CDP-PT:21} and distributed
\cite{AA-JC:20,FC-GB-FP:21,EG-GR:21,JJ-HJVW-HLT-MKC-SH:21} control, as well as
unknown-input
estimation \cite{MST-GFT:21}. We refer the reader to
\cite{IM-FD:21-survey} for a recent survey on data-driven
control.

\smallskip
\noindent
\textbf{Main contributions of this paper.} The main contributions of
this paper are as follows. First, for the linear, discrete,
time-invariant systems described by the triple $(A,B,C)$, we derive
explicit, closed-form data-driven expressions of (i) $\Vstar$, the
largest $(A,\Image(B))$-controlled invariant subspace contained in
$\Ker(C)$, (ii) $\Sstar$, the smallest $(A,\Ker(C))$-conditioned
invariant subspace containing $\Image (B)$, (iii) the feedback gain
$F$ such that $(A+BF) \Vstar \subseteq \Vstar$, and (iv) the invariant 
zeros of $(A,B,C)$.
Since $\Vstar$ and $\Sstar$ are the basis of the geometric approach
developed in \cite{GB-GM:91}, our data-driven formulas constitute the
basis of a data-driven and model-free theory of geometric control, and
can be used to solve a variety of analysis, estimation, and control
problems in a purely data-driven setting. Second, our results show
that the fundamental invariant subspaces of the geometric approach,
which are often computed recursively when operating in the state
space, have a simple and direct interpretation in the
higher-dimensional data space, where they can be computed by solving
appropriately defined sets of linear equations. Third, we demonstrate
the utility of our formulas to design undetectable data-driven
attacks in a consensus system.

\smallskip
\noindent
\textbf{Paper organization.} Section \ref{sec: setup} contains our
problem setup and some preliminary notions. Section \ref{sec:
  data-driven geometric} contains our data-driven formulas of the
fundamental invariant subspaces of the geometric approach. Finally,
Sections \ref{sec: example} and \ref{sec: conclusion} contain our
illustrative examples and conclusion, respectively. 

\smallskip
\noindent
\textbf{Notation.} The set of real numbers is denoted with
$\real$. The rank, range space, null space, transpose, and
Moore-Penrose pseudoinverse of the matrix $A \in \real^{n \times m}$
are denoted with $\rank(A)$, $\Image(A)$, $\Ker(A)$, $A^\top$, and
$A^\dagger$ respectively. A basis of the subspace
$\mc V \subseteq \real^n$ is denoted with $\Basis(\mc V)$.
The Kronecker product between matrices $A$ and $B$ is denoted by $A \otimes B$
and is defined as in
\cite{DSB:09}.

\section{Problem setup and preliminary notions}\label{sec: setup}
We consider the discrete-time linear time-invariant system
\begin{subequations}\label{eq:lti}
  \begin{align} 
    x(t+1) &= Ax(t) + Bu(t) \label{eq:lti_state}\\
    y(t) &= Cx(t) \label{eq:lti_output}
  \end{align}
\end{subequations} 
where $x \in \real^n$, $u \in \real^m$ and $y \in \real^p$ are the
state, input and output vectors, respectively, and $(A,B,C)$ are
constant matrices of appropriate dimensions. For any horizon
$T \ge 1$, the state and output trajectories of \eqref{eq:lti} can be
written as
\begin{equation}\label{eq:X_T}
  \tiny
  \underbrace{
    \begin{bmatrix}
      x(1)\\
      x(2)\\
      \vdots\\
      x(T)
    \end{bmatrix}}_{\subscr{X}{T}}
  =
  \underbrace{
    \begin{bmatrix}
  A \\ 
  A^2 \\
  \vdots \\
  A^{T}
\end{bmatrix}    
  }_{O_T^X} x(0)
 	+ 
 \underbrace{
  \begin{bmatrix}
    B  & \cdots & 0 & 0\\
    AB  & \cdots & 0 & 0 \\
    & \ddots  & & \\
    A^{T-1}B  & \cdots & AB & B
  \end{bmatrix}}_{F_T^X}
                              \underbrace{
                              \begin{bmatrix}
                                u(0)\\
                                u(1)\\
                                \vdots\\
                                u(T-1)
                              \end{bmatrix}}_{\subscr{U}{T}}
                            ,
\end{equation}
and
\begin{equation}\label{eq:Y_T}
  \tiny
  \underbrace{
    \begin{bmatrix}
      y(0)\\
      y(1)\\
      \vdots\\
      y(T-1)
    \end{bmatrix}}_{\subscr{Y}{T}}
  =
  \underbrace{
    \begin{bmatrix}
      C \\ 
      CA \\
      \vdots \\
      CA^{T-1}
    \end{bmatrix}    
  }_{O_T^Y} x(0)
  + 
  \underbrace{
    \begin{bmatrix}
      0 & \cdots & 0 & 0\\
      CB  & \cdots & 0 & 0 \\
      & \ddots  & & \\
      CA^{T-2}B  & \cdots & CB & 0
    \end{bmatrix}}_{F_T^Y}
  \subscr{U}{T}.
\end{equation}
Throughout the paper, we assume that the system matrices are not
known and base our approach on a set of prerecorded
trajectories obtained by arbitrarily probing the system
\eqref{eq:lti}.

\subsection{Data collection}
The available data is collected from a set of $N$ open-loop control
experiments with horizon $T$, and consist of the state and output
trajectories obtained from \eqref{eq:lti} with initial condition
$x^i_0$ and control sequence $U_T^i$, for $i \in \until{N}$. In
particular, the following data matrices are available:
\begin{subequations}\label{eq:data}
  \begin{align} 
    X &= \begin{bmatrix}
      X_T^1 & \!\!\cdots & X_T^N
    \end{bmatrix}  \in \real^{nT \times N} , \label{eq:data_X}
    \\
    X_0 &= \begin{bmatrix} 
      x^1_0 & \,\cdots & x^N_0
    \end{bmatrix}  \;\in \real^{n \times N} , \label{eq : data x0}
    \\
    Y &= \begin{bmatrix}
Y_T^1 & \cdots & Y_T^N
\end{bmatrix}   \in \real^{pT \times N} ,
    \\
    U &= \begin{bmatrix}
      U_T^1 & \cdots & U_T^N
    \end{bmatrix}   \in \real^{mT \times N} . \label{eq:data_U}
  \end{align}
\end{subequations}
From \eqref{eq:X_T}-\eqref{eq:Y_T}, we note the following
relationships:
\begin{equation}\label{eq: x_y_matrix_model}
  \begin{bmatrix}
    X \\
    Y
  \end{bmatrix}
  = \begin{bmatrix}
    O_T^X & F_T^X \\
    O_T^Y & F_T^Y
  \end{bmatrix}
  \begin{bmatrix}
    X_0 \\ 
    U
  \end{bmatrix}.
\end{equation}

We make the following assumption of persistently-exciting experimental
inputs, which is generically satisfied by choosing the inputs and
initial states independently and randomly.
\begin{assumption}\label{ass:excitability}
  The experimental inputs and initial conditions are persistently
  exciting, that is,
  \begin{equation}\label{eq: persistently exciting}
    \rank \begin{bmatrix}
      X_0 \\
      U
    \end{bmatrix} = n + mT.
  \end{equation}
  \oprocend
\end{assumption}

Let $K_0 = \Basis(\Ker(X_0))$ and $K_U = \Basis(\Ker(U))$. From the
Rank-nullity Theorem, Assumption \ref{ass:excitability} ensures that
$X_0 K_U$ and $U K_0$ are full-row rank, respectively. Assumption
\ref{ass:excitability} is a standard assumption in data driven studies
\cite{JC-JL-FD:18,CDP-PT:19}.


\begin{remark}{\emph{\bfseries (Alternative data-driven 
      representations)}}
  Different data formats can be used to obtain a non-parametric
  data-driven representation of the system \eqref{eq:lti}, including
  our representation \eqref{eq:data} as well as Hankel and Page
  matrices \cite{CDP-PT:19,JC-JL-FD:18}. While Hankel and Page
  matrices are generated from a single controlled trajectory, the
  matrices in \eqref{eq:data} use a collection of (possibly shorter)
  controlled trajectories. Different data collections can be more
  convenient for the solution of different problems, with, currently,
  Hankel and Page matrices being used mostly for feedback control
  problems \cite{CDP-PT:19} and multiple trajectories for robustness
  problems \cite{GB-DSB-FP:20,VB-CDP-SF-PT:21}.
\oprocend
\end{remark}

\subsection{Controlled and conditioned invariant subspaces}
The notions of controlled and conditioned invariant subspaces are the
basis of the geometric approach for the analysis and control of linear
systems \cite{GB-GM:69}. We now recall their definition and basic
properties. We refer the interested reader to
\cite{GB-GM:91,WMW:85,HLT-AS-MH:01} for a detailed treatment of this
subject.

\begin{definition}{\emph{\bfseries($(A, \Bc)$-controlled invariant)}}
  Given a matrix $A \in \real^{n \times n}$ and a subspace
  $\Bc \subseteq \real^n$, a subspace $\Vc \subseteq \real^n$ is an
  $(A, \Bc)$-controlled invariant subspace if
  \begin{equation}\label{eq:V}
    A \Vc \subseteq \Vc  + \Bc.
  \end{equation}
\end{definition}
\smallskip

When $\Bc = \Image (B)$, the notion of a controlled invariance refers
to the possibility of confining the state trajectory of the system
\eqref{eq:lti} within a subspace. Specifically, the subspace $\Vc$ is
an $(A, \Image(B))$-controlled invariant subspace if, for every
initial state in $\Vc$, there exists a control input such that the
state belongs to $\Vc$ at all times. Of particular interest is
$\Vstar$, the largest $(A, \Image(B))$-controlled invariant subspace
contained in $\Ker(C)$. The subspace $\Vstar$ contains all
trajectories of \eqref{eq:lti} that generate an identically zero
output. Hence, the subspace $\Vstar$ vanishes if and only if the
system \eqref{eq:lti} features no invariant zeros, a notion that is at
the basis of the analysis of stealthy attacks and unknown-input
observers~\cite{FP-FD-FB:10y}, among others.

\begin{definition}{\emph{\bfseries($(A,\Cc)$-conditioned invariant)}}
  Given a matrix $A \in \real^{n \times n}$ and a subspace
  $\Cc \subseteq \real^n$, a subspace $\Sc \subseteq \real^n$ is an
  $(A, \Cc)$-conditioned invariant subspace if
  \begin{equation}\label{eq:S}
    A(\Sc \cap \Cc) \subseteq \Sc.
  \end{equation}
\end{definition}
\smallskip

When $\Cc = \Ker(C)$, the notion of conditioned invariance arises in
the context of state estimation. Specifically, the subspace $\Sc$ is
an $(A,\Ker(C))$-conditioned invariant subspace if it is possible to
design an (asymptotic) observer that reconstructs the state
$x \setminus \Sc$ by processing the initial condition, the input, and
the measurements of the system \eqref{eq:lti}. Of particular interest is
$\Sstar$, the smallest
$(A,\Ker(C))$-conditioned invariant subspace containing
$\Image(B)$. In fact, the orthogonal complement of the subspace
$\Sstar$ is the largest subspace of the state space that can be
estimated through a dynamic observer in the presence of an unknown
input.

The subspaces $\Vstar$ and $\Sstar$ can be conveniently computed using
simple recursive algorithms \cite{GB-GM:91}. Further, these subspaces
can be used to characterize important properties of the system
\eqref{eq:lti}. For instance, the system \eqref{eq:lti} is right
invertible if and only if $\Vstar \cup \Sstar = \real^n$, and left
invertible if and only if the subspace $\Rstar = \Vstar \cap \Sstar$
is empty \cite{GB-GM:91}. It should be noticed that $\Rstar$
coincides with the largest subspace that can be reached from the
origin with trajectories that belong to $\Vstar$ at all times (hence,
generating an identically zero output).

The definition of the subspaces $\Vstar$, $\Sstar$ and $\Rstar$, as
well as the algorithms to compute them, assume the exact knowledge of
the system matrices. Instead, in the remainder of the paper we derive
purely data-driven expressions of these subspaces, which also offers an
alternative interpretation of them. Similarly to how
$\Vstar$, $\Sstar$ and $\Rstar$ are used in the geometric approach,
our data-driven formulas can also be used to solve a variety of
estimation and control problems.

\begin{figure}[t]
\centering
\begin{tikzpicture} \filldraw[fill=black, opacity=0.1] \bounding;
	\small

    \scope \fill[white] \imaB; \fill[white] \setvstar; \fill[white]
    \setsstar; \fill[white] \setkerC; \fill[white] \setvstari; \endscope

	\begin{scope}
   		\clip \setsstar;
    		\clip \setvstar;
    		\fill[Dandelion, opacity=0.4]\setsstar;
	\end{scope}
	
    \draw \setsstar node [label={[xshift=1.4cm, yshift=1.2cm]$\Sstar$}] {};
    \draw \imaB node [label={[xshift=0.9cm, yshift=-1.6cm]$\Image(B)$}] {};
    \draw \setvstar node [label={[xshift=0cm, yshift=1.4cm]$\Vstar$}] {};
{};
    \draw \setkerC node [label={[xshift=-2.2cm, yshift=1.1cm]$\Ker(C)$}] {};
    \draw \bounding node [label={[xshift=-0.5cm, yshift=-0.8cm]$\real^n$}] {};
	\node at (-0.7,-0.8) {$\Rstar$};
	\node at (-0.6,0)[circle,fill,inner sep=1pt]{};
	\node at (-0.8,0){$0$};

\end{tikzpicture} 
\caption{This figure shows the inclusion relationships between the fundamental
controlled and conditioned invariants defined in Section \ref{sec: setup}. 
Notice how, by their definitions, $\Vstar \subseteq \Ker(C)$, $\Image(B)
\subseteq \Sstar$ and $\Rstar = \Vstar \cap \Sstar$, highlighted in a different
gradient.}
\end{figure}
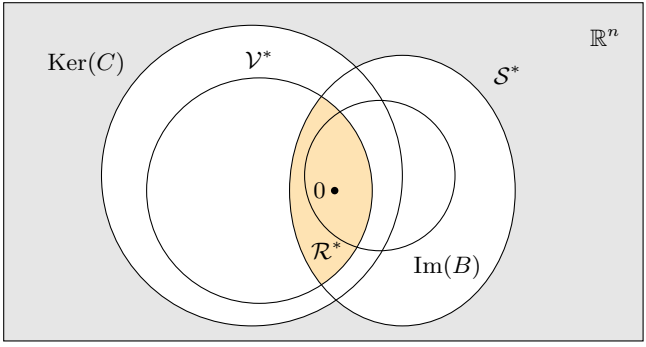

\section{Data-driven geometric control}\label{sec: data-driven geometric}
We begin with finding a data-driven expression of the subspace
$\Vstar$ for the system \eqref{eq:lti}, the largest
$(A,\Image(B))$-controlled invariant subspace contained in
$\Ker(C)$.

\begin{theorem}{\emph{\bfseries (Data driven formula for
      $\Vstar$)}}\label{thm: Vstar}
  Let \eqref{eq:data} be the data generated by the system
  \eqref{eq:lti} with $T \ge n$. Then,
  \begin{equation} \label{eq : Vstar}
    \Vstar =
      \begin{bmatrix}
        X_0 K_U & 0
      \end{bmatrix}
      \Ker
      \begin{bmatrix}
        Y K_U & Y K_0
      \end{bmatrix}
    .
  \end{equation}
  \oprocend
\end{theorem}

To prove Theorem \ref{thm: Vstar}, recall that $\Vstar$ is the set of
initial states for which there exists a control input such that the
resulting state trajectory generates an identically zero output. Since
the system is linear, under our assumption of persistently exciting
experimental inputs, any system trajectory can be expressed as an
appropriate linear combination of the experimental trajectories. We
next formalize this intuition.

\begin{lemma}{\emph{\bfseries (Data-driven trajectories of
      \eqref{eq:lti})}}\label{lemma: trajectory combinations}
  Let \eqref{eq:data} be the data generated by the system
  \eqref{eq:lti} with $T \ge n$. Let $\bar X_T$ and $\bar Y_T$ be the
  state and output trajectories of \eqref{eq:lti} generated with some
  initial condition and control input. Then,
  \begin{equation}\label{eq : x and y bar}
    \begin{bmatrix}
      \bar X_T \\
      \bar Y_T
    \end{bmatrix}
    =
    \begin{bmatrix}
      XK_U & XK_0 \\
      YK_U & YK_0
    \end{bmatrix}
    \begin{bmatrix}
      \alpha\\ 
      \beta
    \end{bmatrix}
    ,
  \end{equation}
  for some vectors $\alpha$ and $\beta$. \oprocend
\end{lemma}
\begin{proof}
  Let $\bar x_0$ and $\bar U_T$ be the initial condition and input to
  \eqref{eq:lti}. Since the matrices $X_0 K_U$ and $U K_0$ are
  full-row rank (see Assumption \ref{ass:excitability}), there exists
  $\alpha$ and $\beta$ such that
  \begin{align} \label{eq: x0 and U from alpha and beta}
    \bar x_0 = X_0 K_U \alpha \;\text{ and }\; \bar U_T = UK_0 \beta .
  \end{align}
  From \eqref{eq:X_T} we have
  \begin{align*}
    \bar X_T &= O_T^X \bar x_0 + F_T^X \bar U_T = O_T^X X_0 K_U
    \alpha+ F_T^X U K_0 \beta \\
    &= X K_U \alpha + X K_0 \beta,
  \end{align*}
  where the last equality follows from \eqref{eq: x_y_matrix_model}. 
  Similarly from \eqref{eq:Y_T},
  \begin{align*}
    \bar Y_T &= O_T^Y \bar x_0 + F_T^Y \bar U_T = O_T^Y X_0 K_U
    \alpha+ F_T^Y U K_0 \beta \\
    &= Y K_U \alpha + Y K_0 \beta,
  \end{align*}
  which concludes the proof.
\end{proof}

Lemma \ref{lemma: trajectory combinations} shows how any state and
output trajectory of \eqref{eq:lti} can be written as a linear
combination of the available data. In particular, state and output
trajectories are obtained in \eqref{eq : x and y bar} as the sum of
the free and forced responses, which are reconstructed from data of
arbitrary control experiments. In fact, $XK_U\alpha$ is the state
trajectory of \eqref{eq:lti} with initial condition $X_0 K_u \alpha$
and zero input (free response), while $XK_0 \beta$ is the state
trajectory of \eqref{eq:lti} with zero initial condition and input
$U K_0 \beta$ (forced response). We remark that Assumption
\ref{ass:excitability} of persistently exciting inputs is necessary
to obtain this result.

The following instrumental Lemma shows that it is sufficient to
consider trajectories of any finite length $T \ge n$ to compute
$\Vstar$, and is instrumental to the proof of Theorem~\ref{thm:
  Vstar}.

\begin{lemma}{\emph{\bfseries (Computing $\Vstar$ from trajectories of
      finite length)}}\label{lemma: finite invisible trajectories}
  For the system \eqref{eq:lti}, any initial state $x_0$, and any
  finite horizon $T \ge n$, the following statements are
  equivalent:
  \begin{enumerate}
  \item $x_0 \in \Vstar$;

  \item there exists an input sequence $u(0),\dots,u(T-1)$ such that
    $y(t) = 0$ for all $t \in \{0,\dots,T-1\}$.
  \end{enumerate}
\end{lemma}
\begin{proof}

  (i) $\Rightarrow$ (ii) Follows from the definition of $\Vstar$.
  
  (ii) $\Rightarrow$ (i) Notice that $y(T-1) = Cx(T-1) = 0$. Thus,
  $x(T-1) \in \Ker(C) = \mathcal V_0$. Similarly, $x(T-2)$ satisfies
  \begin{align*}
    x(T-1) &= Ax(T-2) + Bu(T-2), \text{ and }	\\
    y(T-2) &= Cx(T-2)=0 .
  \end{align*}
  This implies that
  \begin{align*}
    x(T-2) &\in A^{-1} (x(T-1)-Bu(T-2))  \\
    &\subseteq A^{-1}(\mathcal V_0 + \Image(B)) \cap \Ker(C) = \mathcal V_1
  \end{align*}
  Iterating this procedure yields
  \begin{subequations} \label{eq : convergence data driven Vstar}
    \begin{align}
      x(T-1) &\in \mathcal V_0 = \Ker(C), \text{ and } \\
      x(T-i) &\in \mathcal V_{i} = A^{-1}(V_{i-1} + \Image(B)) \cap
               \Ker(C) .
    \end{align}
  \end{subequations}
  Since $\mc V_i$ converges to $\Vstar$ is at most $n$ steps
  \cite{GB-GM:69}, we have that $x(T-\tau) \in \Vstar$ for all
  $\tau \geq n$, which concludes the proof.
\end{proof}

\smallskip
We now prove Theorem \ref{thm: Vstar} using Lemma \ref{lemma:
  trajectory combinations} and \ref{lemma: finite invisible
  trajectories}.


\begin{pfof}{Theorem \ref{thm: Vstar}} From Lemma \ref{lemma: finite
    invisible trajectories} we seek all initial conditions $x_0$ for
  which the output can be maintained at zero for $T \ge n$ steps. From
  \eqref{eq : x and y bar}, the vectors $\alpha$ and $\beta$ that
  identify state trajectories with identically zero output must
  satisfy
\begin{equation}\label{eq : alpha and beta for Vstar}
  \begin{bmatrix}
    \alpha 
    \\
    \beta
  \end{bmatrix}
  \in \Ker 
  \begin{bmatrix}YK_U & YK_0 \end{bmatrix}.
\end{equation}
The initial condition corresponding to such trajectories is
$x_0 = X_0 K_U \alpha$ (see \eqref{eq: x0 and U from alpha and
  beta}). Thus, the set $\Vstar$ can be written as
\[
  \Vstar = 
  \begin{bmatrix}
    X_0 K_U & 0
  \end{bmatrix}
              \Ker
              \begin{bmatrix}
                Y K_U & Y K_0
              \end{bmatrix}.
\]
\hfill\QEDA
\end{pfof}


We next find a data-driven expression for $\Sstar$, the smallest
$(A,\Ker(C))$ conditioned invariant containing $\Image(B)$.

\begin{theorem}{\emph{\bfseries (Data driven formula for
      $\Sstar$)}}\label{thm: Sstar}
  Let \eqref{eq:data} be the data generated by the system
  \eqref{eq:lti} with $T \ge n$. Then,
  \begin{equation} \label{eq : Sstar}
    \Sstar =
    H X K_0
    \Ker \left(Y K_0 \right)
    ,
  \end{equation}
  where\footnote{We use $I_n\in \real^{n \times n}$ and
  $0_{n} \in \real^{n \times n}$ to denote the identity and
  zero matrices of appropriate dimensions, respectively.} $H=
\underbrace{\begin{bmatrix}
      0_{n} & \cdots & 0_{n} & I_{n}
    \end{bmatrix}}_{T\text{ matrices}}$.
\end{theorem}

\smallskip
To prove Theorem \ref{thm: Sstar}, we first show that, similarly to
the case of $\Vstar$, the subspace $\Sstar$ can be computed from a
collection of trajectories of finite length $T \ge n$.

\begin{lemma}{\emph{\bfseries (Computing $\Sstar$ from trajectories of
      finite length)}} \label{lemma: finite invisible trajectories S}
  For the system \eqref{eq:lti} and any finite horizon $T \geq n$, the
  following statements are equivalent:
  \begin{enumerate}
  \item $x(T) \in \Sstar$;

  \item there exists an input sequence $u(0),\dots,u(T-1)$ such that
    $y(t) = 0$ for all $t \in \{0,\dots,T-1\}$ and $x(0) = 0$.
  \end{enumerate}
\end{lemma}
\begin{proof}

  (i) $\Rightarrow$ (ii) Follows from the definition of $\Sstar$. For example,
  with $u(t) = 0$, for $t \in \fromto{0}{T-2}$ and $u(T-1)
  \neq 0$. Then $x(T) = Ax(T-1) + Bu(T-1) = Bu(t-1) \in \Image(B) \subseteq
  \Sstar$. 
  
  \smallskip
  (ii) $\Rightarrow$ (i) Because $x(1) = B u(0)$ and $y(1) = C x(1) = 0$, we
  have $x(1) \in \Image(B) \cap \Ker(C) = \Sc_1 \cap \Ker(C)$. Similarly,
  \begin{align*}
    x(2) &\in A(\Sc_1 \cap \Ker(C)) + \Image(B) = \Sc_2,
  \end{align*}
  and $x(2) \in \Ker(C)$ since $y(2) = C x(2) = 0$. Recursively:
  \begin{subequations} \label{eq : convergence data driven Sstar}
    \begin{align}
      x(1) &\in \Sc_1 = \Image(B), \text{ and } \\
      x(i) &\in \Sc_{i} = A(\Sc_{i-1} \cap\Ker(C)) + \Image(B) .
    \end{align}
  \end{subequations}
  Since $\mc S_i$ converges to $\Sstar$ in at most $n$ steps
  \cite{GB-GM:69}, we have that $x(\tau) \in \Sstar$ for all $\tau \ge
  n$, which concludes the proof.
\end{proof}

\smallskip

We are now ready to prove Theorem \ref{thm: Sstar}.

\begin{pfof}{Theorem \ref{thm: Sstar}}
  From \eqref{eq : x and y bar}, when $x(0) = 0$, any state trajectory
  of length $T$ that generates an identically zero output of length
  $T$ can be parametrized with $\alpha = 0$ and
  $\beta \in \Ker(Y K_0)$. Using Lemma \ref{lemma: finite invisible
    trajectories S}, the set $\Sstar$ can be equivalently written as
  the final states reached by such trajectories, that is,
  $\Sstar = H X K_0 \Ker(Y K_0)$, which concludes the proof.
  \hfill\QEDA
\end{pfof}

\begin{remark} {\emph{\bfseries (Obtaining $\Rstar$ from $\Vstar$ and
$\Sstar$)}}
The combined knowledge of $\Vstar$ and $\Sstar$ allows us to find
$\Rstar$ as \cite{GB-GM:91}
\begin{equation} \label{eq : Rstar intersection}
	\Rstar = \Vstar \cap \Sstar.
\end{equation}
Alternatively, one can also find explicit data-driven expression for
$\Rstar$ using trajectories of finite-length. For example, one can
show that when the condition $u(0) \neq 0$ is imposed in Lemma
\ref{lemma: finite invisible trajectories S}, then
$x(T) \in \Rstar \subseteq \Sstar$. We
omit the proof of this result, and use \eqref{eq : Rstar intersection}
to directly compute $\Rstar$. \oprocend
\end{remark}

\smallskip The state of a system can be confined within a subspace
$\mc V$ through a state-feedback controller if and only if $\mc V$ is
a controlled invariant subspace. We continue this section with the
data-driven design of such state-feedback controller, that is, the
data-driven design of a matrix $F$ such that
\begin{align}\label{eq:F}
  (A+BF)\Vc \subseteq \Vc.
\end{align} 
For a trajectory $X_T$ and input $U_T$, let
\begin{subequations}\label{eq:data_cl}
\begin{align}
  X_{0,T} &=
  \begin{bmatrix}
    x(1) & x(2) & \cdots & x(T-1)
  \end{bmatrix} , \\
  X_{1,T} &=
  \begin{bmatrix}
    x(2) & x(3) & \cdots & x(T)
  \end{bmatrix}                           
                             , \text{ and }\\
  U_{0,T} &=
  \begin{bmatrix}
    u(0) & u(1) & \cdots & u(T-1)
  \end{bmatrix}
                           .
\end{align}
\end{subequations}

\begin{theorem}{\emph{\bfseries (Data-driven feedback for invariant
      subspace)}}\label{thm:data driven F}
  Let $X_T$ be the trajectory of \eqref{eq:lti} with input $U_T$ and
  some initial condition. Let $\Vc = \Image(V)$ be an
  $(A,\Image(B))$-controlled invariant subspace, and let
  \begin{equation}\label{eq:Fdd}
    F = U_{0,T}(X_{0,T}^\dag + K \gamma) ,
  \end{equation}
  with $K = \Ker(X_{0,T})$ and
  \begin{equation} \label{eq:gamma} \gamma = - ((I - V V^\dag)X_{1,T}
    K)^\dag (I - V V^\dag) X_{1,T} X_{0,T}^\dag V V^\dag.
  \end{equation}
  If $[U_{0,T}^\top \; X_{0,T}^\top]^\top$ is full row
  rank,\footnote{This condition requires the trajectory to be
    sufficiently informative and is related to the notion of
    persistency of excitation
    \cite{JCW-PR-IM-BLMDM:05,JC-JL-FD:18,CDP-PT:19}.} then
  $(A+BF)\Vc \subseteq \Vc$.
\end{theorem}
\begin{proof}
  From \cite[Theorem~2]{CDP-PT:19}, for any state-feedback gain
  $F$, the closed loop matrix can be written as
  \begin{align*}
    A+BF = X_{1,T} G,
  \end{align*}
  where the matrix $G$ satisfies $X_{0,T} G = I$ and $U_{0,T} G =
  F$. Further, $F$ renders the subspace $\mc V$ invariant if and only
  if
  \begin{align*}
    (A+BF)\mc V = X_{1,T} G \mc V \subseteq \mc V,
  \end{align*}
  or, equivalently,
  \begin{align*}
    (I - V V^\dag) X_{1,T} G V = 0,
  \end{align*}
  where $V = \Basis(\mc V)$ and $(I - V V^\dag)$ is a projector onto
  $\mc V^\perp$. From $X_{0,T} G = I$ we obtain
  $G = X_{0,T}^\dag + K \gamma$, where $\gamma$ is any matrix of
  \begin{align*}
    (I - V V^\dag) X_{1,T} (X_{0,T}^\dag + K \gamma) V = 0.
  \end{align*}
  Solving for $\gamma$ (a solution $\gamma$ exists because
  $\mc V$ is an $(A,B)$-controlled invariant subspace and
  $[U_{0,T}^\top \; X_{0,T}^\top]^\top$ is full-row rank) and using
  $U_{0,T} G = F$ concludes the proof.
\end{proof}

Theorem \ref{thm:data driven F} details the computation of a feedback
matrix that renders a subspace invariant, from sufficiently
informative state and input trajectories. It should be noticed that
Theorem \ref{thm:data driven F} does not guarantee the internal, nor
external, stability of the subspace, which imposes additional
constraints on $\gamma$. This is left as a topic of future
investigation.

To conclude this section we present a strategy to identify the
invariant zeros of \eqref{eq:lti} from data. We make the assumption
that \eqref{eq:lti} is \emph{non-degenerate}, i.e., $\Rstar$ is empty.
Degenerate systems are
intrinsically vulnerable to, e.g., undetectable malicious
attacks with unstable state trajectories. On the other hand, for non-degenerate
systems, the existence of unstable invisible trajectories depends on the
modulo of its invariant zeros. In fact, the knowledge of
the number and magnitude of the invariant zeros of a non-degenerate
system is essential when studying problems such as noninteracting control
\cite{GB-GM:69} and malicious attack detection \cite{FP-FD-FB:10y}, motivating
our interest in their identification.  


\begin{theorem}{\emph{\bfseries (Data-driven invariant zeros)}} \label{thm:
zeros}
  Let $X$ and $\Vstar$ be as in \eqref{eq:data_X} and \eqref{eq :
    Vstar}, respectively, with $T \geq n$.  Let $V = \Image(\Vstar)$
  and assume that $\Rstar = \emptyset$.  Then, $z \in \complex$ is an
  invariant zero of \eqref{eq:lti} if and only if the matrix
  \begin{equation}
    \left[X X^\dag (I \otimes V)
      \quad -\left([z ~ z^2 ~ \cdots ~ z^T] \otimes I\right)^\top \right] 
      \end{equation}
           has a nontrivial kernel.
\end{theorem}
\smallskip
\begin{proof}
  When $\Vstar \neq \emptyset$ and $\Rstar = \emptyset$,
  there exists a trajectory $x(t) = z^t x(0)$, with $x(t) \in \Vstar$
  for all $t \geq 0$ and $z$ an invariant zero of \eqref{eq:lti}
  \cite{GB-GM:91}. We write such trajectory as
  \begin{equation}\label{eq: X_T^V_1}
    X_T^V =
    \begin{bmatrix}
      zI \\
      z^2I\\
      \vdots \\
      z^T I
\end{bmatrix} v = \left([z ~ z^2 ~ \cdots ~ z^T] \otimes
I
    \right)^\top v.
  \end{equation}
  With Assumption \ref{ass:excitability}, any trajectory belongs to
  the image of the data matrix $X$. Then, when the trajectory $X_T^V$
  above exists, there also exists a vector $\bar w \in X^\dag (I \otimes V)$
  such that $X\bar w = X_T^V$. The condition on $\bar w$ imposes that the
  trajectory
  is compatible with \eqref{eq:lti} while evolving inside $\Vstar$. Both
  vectors $v \neq 0$ and $\bar w = X^\dag (I \otimes V)w \neq 0$ exist
  if and only if 
\begin{equation}
	X X^\dag (I \otimes V) w = \left([z ~ z^2 ~ \cdots ~ z^T] \otimes
I\right)^\top v
\end{equation}
i.e., the kernel of
$[X X^\dag (I \otimes V)
  \quad -\left([z ~ z^2 ~ \cdots ~ z^T] \otimes I\right)^\top]$ is 
  non-empty, concluding the proof. 
\end{proof}

The invariant zeros of the system \eqref{eq:lti} can be equivalently
characterized using data collected as in \eqref{eq:data_cl}.
\begin{lemma}{\emph{\bfseries (Data-driven invariant zeros)}}
  Let $\Vstar$ be as in \eqref{eq : Vstar} and assume that $\Rstar =
\emptyset$.
Let $T =
  \begin{bmatrix}
    T_1 & T_2
  \end{bmatrix},$
  with $T_1 = \Vstar$, and $T_2$ chosen such that $T$ is
  nonsingular. 
  Finally, let $G=X_{0,T}^\dag + K \gamma$, with $\gamma$
  defined as in \eqref{eq:gamma}.
  Then, the invariant zeros of \eqref{eq:lti} are the
  eigenvalues of $A_{11}$, where
  \begin{equation}\label{eq:Aprime}
    T^{-1}(X_{1,T}G) T = \begin{bmatrix}
      A_{11} & A_{12}  \\
      0 & A_{22}  \\
    \end{bmatrix}.
  \end{equation}
\end{lemma}
\smallskip
\begin{proof}
  This result derives from the facts that (i) the closed loop system
  with the state feedback $u = Fx$ satisfies
  \begin{equation}
    A + BF = X_{1,T}G,
  \end{equation}
  (ii) the subspace $\Vstar$ is invariant for the closed-loop matrix
  $A + BF$, and (iii) the invariant zeros of \eqref{eq:lti} are the
  eigenvalues of the closed-loop matrix $A + BF$ contained in
  $\Vstar$.
\end{proof}

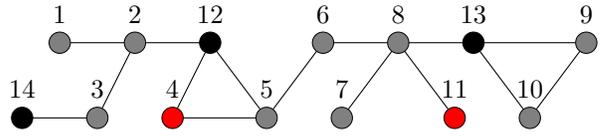
\begin{figure}[t]
	\centering
	\begin{tikzpicture}
		\node[fill=gray,circle,draw,ultra thin,inner sep=0.1cm,label=above:$1$]
				(1) at (0,0) {};
		\node[fill=gray,circle,draw,ultra thin,inner sep=0.1cm,label=above:$2$] 
		        (2) at (1,0) {};
		\node[fill=gray,circle,draw,ultra thin,inner sep=0.1cm,label=above:$3$] 
				(3) at (0.5,-1) {};
		\node[fill=black,circle,draw,ultra thin,inner sep=0.1cm,label=above:$14$] 
				(14) at (-0.5,-1) {};
		\node[fill=black,circle,draw,ultra thin,inner sep=0.1cm,label=above:$12$]
				(12) at (2,0) {};
		\node[fill=red,circle,draw,ultra thin,inner sep=0.1cm,label=above:$4$]
				(4) at (1.5,-1) {};
		\node[fill=gray,circle,draw,ultra thin,inner sep=0.1cm,label=above:$5$] 
				(5) at (2.75,-1) {};
		\node[fill=gray,circle,draw,ultra thin,inner sep=0.1cm,label=above:$6$] 
				(6) at (3.5,0) {};
		\node[fill=gray,circle,draw,ultra thin,inner sep=0.1cm,label=above:$8$] 
				(8) at (4.5,0) {};
		\node[fill=gray,circle,draw,ultra thin,inner sep=0.1cm,label=above:$7$] 
				(7) at (3.75,-1) {};
		\node[fill=red,circle,draw,ultra thin,inner sep=0.1cm,label=above:$11$]
				(11) at (5.25,-1) {};
		\node[fill=black,circle,draw,ultra thin,inner sep=0.1cm,label=above:$13$] 
				(13) at (5.5,0) {};
		\node[fill=gray,circle,draw,ultra thin,inner sep=0.1cm,label=above:$9$]
		 		(9) at (7,0) {};
		\node[fill=gray,circle,draw,ultra thin,inner sep=0.1cm,label=above:$10$] 
				(10) at (6.25,-1) {};
				
		\draw (1) to (2);
		\draw (2) to (3);
		\draw (3) to (14);
		\draw (2) to (12);
		\draw (12) to (4);
		\draw (12) to (5);
		\draw (4) to (5);
		\draw (5) to (6);
		\draw (6) to (8);
		\draw (8) to (7);
		\draw (8) to (11);
		\draw (8) to (13);
		\draw (13) to (9);
		\draw (13) to (10);
		\draw (9) to (10);
	\end{tikzpicture}
	\caption{An example of consensus network from \cite{ME-SM-MC-KC-AB:12}. 
	Agents are numbered from $1$ through $14$, 
	where nodes $\{12,13,14\}$ (in black) are the leaders and nodes
	$\{4,11\}$ (in red) are the network monitors.}
	\label{fig:network}
      \end{figure}
      
\section{Malicious attacks: an illustrative example}
\begin{figure}[t]
  \centering
  \includegraphics[width=0.99\columnwidth]{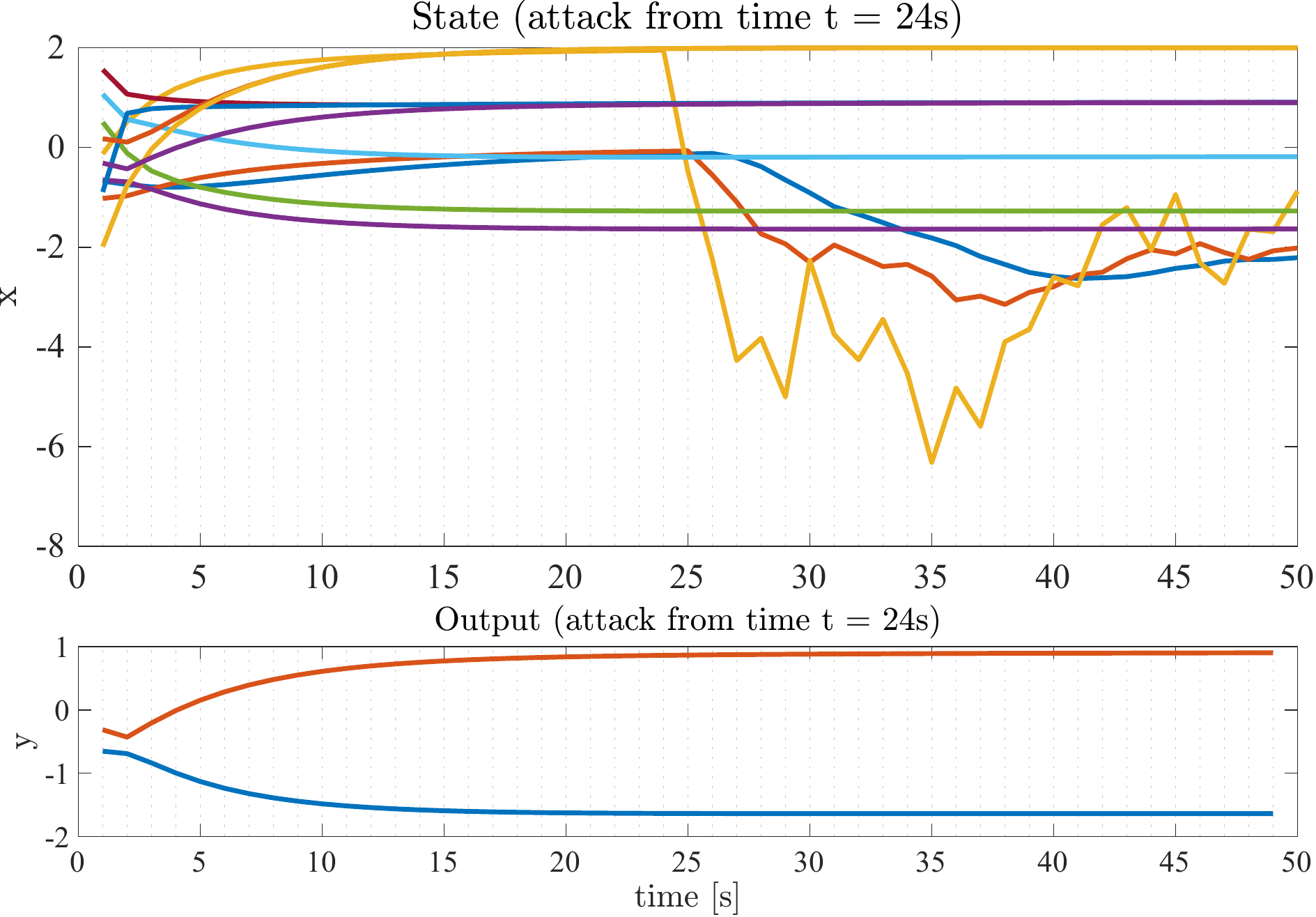}
  \caption{In this figure we show an attack on the network of Fig.
    \ref{fig:network}. The systems initial condition is chosen
    randomly and the leaders impose $u = [-2 ~ 2 ~ 4]^\top$.  The
    attacker waits for the system to reach its equilibrium and then,
    at time $t = 24s$, injects an attack $A_T$ as proposed in Sec
    \ref{sec: example}.  We notice how the system state its perturbed
    from the equilibrium, while the output of the system remains
    unaffected by the attack, rendering the attack action effectively
    invisible at the output.}
  \label{fig:attack}
\end{figure}

\label{sec:
  example}
To illustrate a possible use of the theory developed in this paper,
consider the leader-follower consensus network in
Fig. \ref{fig:network}. The dynamics of the followers are given by the
matrices
\begin{align*}
  A &= \begin{bmatrix}
    .8 & .2 & 0 & 0 & 0 & 0 & 0 & 0 & 0 & 0 & 0 \\
    .2 & .4 & .2 & 0 & 0 & 0 & 0 & 0 & 0 & 0 & 0 \\
    0 & .2 & .6 & 0 & 0 & 0 & 0 & 0 & 0 & 0 & 0 \\
    0 & 0 & 0 & .6 & .2 & 0 & 0 & 0 & 0 & 0 & 0 \\
    0 & 0 & 0 & .2 & .4 & .2 & 0 & 0 & 0 & 0 & 0 \\
    0 & 0 & 0 & 0 & .2 & .6 & 0 & .2 & 0 & 0 & 0 \\
    0 & 0 & 0 & 0 & 0 & 0 & .8 & .2 & 0 & 0 & 0 \\
    0 & 0 & 0 & 0 & 0 & .2 & .2 & .2 & 0 & 0 & .2 \\
    0 & 0 & 0 & 0 & 0 & 0 & 0 & 0 & .6 & .2 & 0 \\
    0 & 0 & 0 & 0 & 0 & 0 & 0 & 0 & .2 & .6 & 0 \\
    0 & 0 & 0 & 0 & 0 & 0 & 0 & .2 & 0 & 0 & .8
  \end{bmatrix}
                                             ,\\
  B^\top &= \begin{bmatrix}
    0 & .2 & 0 & .2 & .2 & 0 & 0 & 0 & 0 & 0 & 0 \\
    0 & 0 & 0 & 0 & 0 & 0 & 0 & .2 & .2 & .2 & 0 \\
    0 & 0 & .2 & 0 & 0 & 0 & 0 & 0 & 0 & 0 & 0
  \end{bmatrix}
                                             ,
                                             \text{ and }\\
  C &= \begin{bmatrix}
    0 & 0 & 0 & 1 & 0 & 0 & 0 & 0 & 0 & 0 & 0 \\
    0 & 0 & 0 & 0 & 0 & 0 & 0 & 0 & 0 & 0 & 1
  \end{bmatrix}
                                            .
\end{align*}
The network is equipped with two monitoring nodes, specifically, nodes
$4$ and $11$. The state of the monitoring nodes is used to detect any
anomalous behavior of the network from its nominal dynamics (see also
\cite{FP-AB-FB:09b}). We let an attacker take control of the leader
nodes, and seek for an attack strategy that remains undetectable from
the monitoring nodes, and leverages only historical data of the
network dynamics. In particular, the attacker strategy is designed as
follows: (i) compute $\Vstar$ and $\Sstar$ using Theorems \ref{thm:
  Vstar} and \ref{thm: Sstar}, respectively, and find
$\Rstar = \Vstar \cap \Sstar$; (ii) for $R = \Basis(\Rstar)$, and $X$,
$U$ and $K_0$ defined as in \eqref{eq:data_X}, \eqref{eq:data_U} and
Assumption \ref{ass:excitability}, compute\footnote{Similarly to the
  proof of Theorem \ref{thm: zeros}, it can be shown that, for
  nontrivial $v$ and $w$, any trajectory satisfying
  $XK_0 v = (I \otimes R)w$ (i) starts at the origin, (ii) evolves in
  $\Rstar$, and (iii) is compatible with the data \eqref{eq:data} of
  \eqref{eq:lti}.}  $P$ as
\begin{equation}
  \begin{bmatrix}
		XK_0 & I \otimes R
	\end{bmatrix} \begin{bmatrix}
		P \\ Q
	\end{bmatrix} = 0 ;
\end{equation}
and (iii) choose the attack input $A_T$ as $A_T \in
\Image(UK_0P)$. Then, for any initial state $x(0)$ and nominal control
input $U_T$, the output of \eqref{eq:X_T}-\eqref{eq:Y_T} with input
$U_T$ is indistinguishable from the output with input $U_T + A_T$.  As
can be seen in Fig. \ref{fig:attack} from time $t = 24s$, the attacker
strategy perturbs the state of the network but does not affect the
monitoring nodes, thus remaining undetectable. In fact, it can be
shown that any input $A_T \in \Image(UK_0P)$ moves the state
trajectory within the controlled invariant $\Rstar \subseteq \Ker(C)$,
thus affecting the state of the system but not its output.

\section{Conclusion}\label{sec: conclusion}
In this paper we show how experimental data can be used to learn key
invariant subspaces of a linear system. In particular, we derive
data-driven expressions for $\Vstar$, the largest
$(A,\Image(B))$-controlled invariant contained in $\Ker(C)$, and
$\Sstar$, the smallest $(A,\Ker(C))$-conditioned invariant containing
$\Image(B)$. Being able to identify these subspaces from data suggests
that much of the results and intuitions of the geometric approach to
control can be conveniently reworked in a data-driven framework.  To
support this point, we leverage the identified invariant subspaces to
design a data-driven feedback controller to force the state inside a
desired controlled invariant subspace, and to compute the invariant
zeros of the system. Finally, as an example of the theoretical
results, we design a data-driven undetectable attack. Applications and
extensions of the proposed results are numerous, and are left as the
subject of future investigation.



\bibliographystyle{ieeetr}
\bibliography{alias,FP,Main,New}
\end{document}